\documentclass[final,3p,times,twocolumn]{elsarticle}
\usepackage{graphicx}
\usepackage{subfigure}
\usepackage{indentfirst}
\usepackage{amsmath}
\usepackage{caption2}
\usepackage{algorithm}
\usepackage{algorithmic}
\usepackage{amsthm}
\usepackage{multirow}
\journal{Computer Communications}

\begin{document}
  \begin{frontmatter}
\title{Phase Changes in the Evolution of the IPv4 and IPv6 AS-Level Internet Topologies}
\author{Guoqiang Zhang}
\address{Institute of Computing Technology, Chinese Academy of Sciences, Beijing, China\\ Email: guoqiang@ict.ac.cn}
\author{Bruno Quoitin}
\address{Networking Lab, Universit$\acute{e}$ de Mons, Place du Parc, 20, 7000, Mons, Belgium}
\author{Shi Zhou}
\address{Department of Computer Science, University College London, United Kingdom}

 \begin{abstract}


In this paper we investigate the evolution of the IPv4 and IPv6 Internet topologies at the autonomous system
(AS) level over a long period of time.
We provide abundant empirical evidence that there is a phase transition in the growth trend of the two
networks.
For the IPv4 network, the phase change occurred in 2001.
Before then the network's size grew exponentially, and thereafter it followed a linear growth. Changes are
also observed around the same time for the maximum node degree, the average node degree  and the average
shortest path length.
For the IPv6 network, the phase change occurred in late 2006.
%
It is notable that the observed phase transitions in the two
networks are different, for example the size of IPv6 network
initially grew linearly and then shifted to an exponential growth.
Our results show that following decades of rapid expansion up to the
beginning of this century, the IPv4 network has now evolved into a
mature, steady stage characterised by a relatively slow growth with
a stable network structure; whereas the IPv6 network, after a slow
startup process, has just taken off to a full speed growth. We also
provide insight into the possible impact of IPv6-over-IPv4 tunneling
deployment scheme on the evolution of the IPv6 network.
The Internet topology generators so far are based on an inexplicit assumption that the
evolution of Internet follows non-changing dynamic mechanisms. This assumption, however, is invalidated by
our results.
Our work reveals insights into the Internet evolution and provides
inputs to future AS-Level Internet models.

 \end{abstract}

\end{frontmatter}

\section{Introduction}

The Internet has experienced rapid growth in the past 30 years, evolving from a simple laboratory test-bed
network to a gigantic ecosystem. It is often considered as the most complex technological network ever made
by human beings. From the highest level, this ecosystem can be represented by a graph, where nodes represent
the autonomous systems (ASes), and two nodes are connected if and only if the two ASes are engaged in a
business relationship to exchange data traffic.

Since late 1990s', various research activities are devoted to the
mapping, characterisation and modelling of the Internet
\cite{power-law-99,densification-laws,three-data-sources,collecting-as-topology,representative-as-topology,AS-path-inflation,
symbiotic-effect, CAIDA,DIMES,Routeviews,ripe,
in-search-for-appropriate,building-as-topology-model}. These efforts
have indeed uncovered intriguing features of the Internet, e.g.,
power-law degree distribution \cite{power-law-99}, rich-club
phenomenon \cite{rich-club}, disassortative mixing \cite{mixing},
self-similarity \cite{chinese-as-graph}, etc. These discoveries are
further followed by proposals of different network models that try
to reproduce these distinctive topological properties
\cite{BA,AB,GLP,PFP,understanding-the-evolution-of-internet,network-topology-generators}.
Readers can refer to \cite{network-topologies-survey} for a survey
of network modeling and generation.

However, despite the significant amount of efforts, existing studies
still face several problems and challenges:
\begin{itemize}
\item Firstly, although tremendous Internet measurement projects are set up,
we still cannot have a comprehensive and accurate view of the real
AS topology \cite{impacts-sampling-bias,
internet-dark-matter,a-framework-for-discovering-missing-links,in-search-of-the-elusive-ground-truth}.
This is because the AS topology inference methods, either BGP-based
or traceroute-based, suffer a common problem of systematic loss of a
nontrivial fraction of links, mostly peer-to-peer links between
periphery nodes.
\item Secondly, most studies are carried out
on particular \emph{snapshots} of the Internet topology or over short-term historic data (less than 5 years),
e.g., topological properties are uncovered for particular snapshots, and network models are validated by
particular observed snapshots. Relatively few efforts have been put to the evolutionary study of the Internet
topology over a long time period.
\item Thirdly, of the limited number of studies on the evolution of
the Internet, researchers often do not determine the real causes for observed topology changes. Some of the
changes may not due to real evolution events but originate from the variation of monitors
\cite{densification-laws,graph-mining,ten-years-evolution,evolution-of-internet-and-core}. This makes their
claims questionable.
\item Finally, the Internet
now is  experiencing a gradual transition from the IPv4 network to the IPv6 network due to a number of
reasons including the shortage of IP addresses. A natural question is whether these two networks show similar
or different evolutionary trends. Yet, to the best of our knowledge, very few work has been done to  study
the evolution of the IPv6 network, let alone a side-by-side study of the two networks.
Without this study, problems such as how the IPv6-over-IPv4
tunneling impacts on the evolution of the IPv6 network could not be
properly understood.
\end{itemize}

Motivated by these, in this paper we undertook an in-depth side-by-side study of the evolution of the IPv4
and IPv6 AS-Level Internet topologies over a long period of time.
%
%
We aim to answer questions such as: whether the Internet has a
uniform evolution process, or experiences different evolution
stages? whether its featured structural properties keep unchanged,
or evolve over time? and whether the existing network models are
capable of modeling the real evolution process of the Internet?
More specifically, our original contributions are:
\begin{enumerate}


\item We are the first to carry out a long-term side-by-side
evolutionary study of the IPv4 and IPv6 network topologies at the AS
level.

\item Based on historic routing data, we show amble empirical evidence that both the IPv4 and IPv6 networks have experienced a phase change in their evolution,
but with different transition patterns. The IPv4 network has evolved
into a stable structure, whereas the IPv6 network has just entered a
stage of rapid growth. Notably, it is the first time in the
literature to discover phase change in the evolution of the IPv6
network.

\item We have discussed the impact of IPv6-over-IPv4 tunneling deployment scheme on the
evolution of the IPv6 network.

\item We point out the fundamental impact of the  phase changes of the
Internet evolution on designing and evaluating future Internet models.
\end{enumerate}

The following of the paper is organised as such. Related work is discussed in Section \ref{related-work}.
Section \ref{data-set} presents the data sets and approaches we use for this study. Section
\ref{evolution-study} gives the side-by-side evolution study of  the IPv4 and IPv6 AS-Level topologies. We
 discuss our findings in Section \ref{impacts}. Finally, we
conclude the paper in Section \ref{conclusion}.

\section{Related Work}\label{related-work}
The last decade has witnessed a surge of research activities related
to network topology measurement, characterization and modeling.
Various projects are set up to map the Internet topology. The BGP
table dumps archived by Routeviews \cite{Routeviews} and RIPE
\cite{ripe} offer good feeds for the study of AS-Level Internet
topology. The outcome of the active measurement projects, such as
CAIDA \cite{CAIDA} and DIMES \cite{DIMES}, on the other hand,
provides input to studies for both the AS-level and router-level
Internet topologies.

These data sources provide researchers with an unprecedented
opportunity to uncover the unique structural properties as well as
evolutionary mechanisms of this complex man-made system. Various
topological features are discovered for specific topology snapshots,
e.g., power-law degree distribution \cite{power-law-99}, assortative
mixing \cite{mixing}, rich-club phenomenon \cite{rich-club},
extremely large maximum degree
\cite{PFP,evolution-of-internet-and-core}, high clustering
coefficient \cite{newman-review}, and self-similarity between
regional AS subgraph and the global AS graph\cite{chinese-as-graph}.
These analysis were followed by a number of graph theory based
generative models to reproduce the observed characteristics and try
to explain the evolution of networks, e.g., BA \cite{BA}, AB
\cite{AB}, GLP \cite{GLP}, PFP
\cite{PFP,understanding-the-evolution-of-internet}.

Recently, there is a growing trend to study the Internet from an evolutionary perspective. Based on the early
day's data from Routeviews, it was shown in \cite{densification-laws} and \cite{graph-mining} that the
AS-Level Internet topology was densifying and its effective diameter was shrinking. In
\cite{observing-internet-evolution}, the authors grouped the BGP data from Routeviews and RIPE into three
sets to evaluate the effects of different monitors, i.e., data from a single monitor, data from a fixed
number of monitors that are present throughout the entire measurement period, and data from all monitors. It
was shown that after a short exponential revealing period, the network follows a constant birth rate. In
\cite{ten-years-evolution}, the authors carried out a ten-year study of the evolution of the Internet(the
longest time period among this kind of studies before this paper) and it was shown that the number of ASes as
well as the number of CP(customer-provider) links follow similar growth trends, that is, both grow
exponentially from Nov, 1997 to May, 2001, and then enter into a linear growth mode. In our recent work
\cite{evolution-of-internet-and-core}, we reported that the maximum degree remains nearly invariable in
recent years, and the so-called $k$-core property is stable over time.


Only recently, there has been some effort towards characterising and
modeling the IPv6 network. CAIDA's Ark project began to perform
continuous large-scale active measurement of the IPv6 network since
Dec, 2008 \cite{visualizing-ipv6}. In \cite{modeling-ipv6}, it was
shown that although the IPv6 AS topology obeys power law, its degree
exponent is much smaller than the IPv4 counterpart, and a novel
model was proposed to reproduce this smaller degree exponent
phenomenon. Wesley M. Eddy~\cite{basic-property-ipv6} took a
three-year-long evolutionary study of the IPv6 AS-level topology.
However, since the author only studied the data from May, 2003 to
Sep, 2006, he did not observe the phase transition that took place
in 2006 as we will report in the following.

\section{The data set}\label{data-set}
In this study, we used the data set offered by Routeviews and RIPE
since they are the only public sources that archive historic BGP
data. We do not use the AS topological data derived from traceroute
measurements due to issues in converting router paths to AS paths
\cite{observing-internet-evolution,third-party-address-in-traceroute-paths,towards-an-accurate-traceroute-tool}.
We used an approach similar to \cite{observing-internet-evolution}
to group the data into different sets to evaluate the effects of
different monitors. For IPv4, we built three different data sets:
\begin{itemize}
\item {\bf OIX:} data from the single Routeviews collector \emph{route-views.route-views.org},
which is extensively used in early day's AS topology analysis. We
collected the data from the starting date of the collector, i.e.,
Nov, 1997.
\item {\bf Set52:} data from a set of 52 monitors in both Routeviews and RIPE. These 52 monitors reside in 36 ASes that persist all the time since
Jul, 2004. According to the AS taxonomy provided by CAIDA
\cite{as-taxonomy, revealing-as-taxonomy}, the 36 ASes contain 11
tier-1 ASes, 19 tier-2 ASes, 2 NICs, and 4 abstained
ASes\footnote{abstained means the
algorithm~\cite{revealing-as-taxonomy} fails to make a predication
of the AS class.}.
\item {\bf ALL:} data from all the collectors of Routeviews (except route-views6) and RIPE
that started prior to Jul, 2004.
\end{itemize}

For IPv6, we built two data sets:
\begin{itemize}
\item {\bf Set4:} data from four monitors (residing in the following four ASes: AS2497, AS2914, AS7660, AS30071) that occurred most frequently
since May, 2003, with each occurring more than 60 times out of the
77 months.
\item {\bf ALL: } data from the route-views6 collectors since
May, 2003.
\end{itemize}

We collect the data on a monthly basis. Each month, we collected one snapshot from each collector in the last
day of the month with collection time as close as possible for different monitors, and then synthesised the
AS paths from different monitors to construct the corresponding data set. AS paths that contain AS set,
private ASNs, or loops were filtered out from the graph construction. Although collecting only one snapshot
in a month can miss some hidden links that could be revealed at a later time, merging all the snapshots over
a relatively long time period, however, can potentially introduce the problem of stale links
\cite{a-framework-for-discovering-missing-links,in-search-of-the-elusive-ground-truth,observing-internet-evolution}.
We thus focused on an \emph{instant operating view} of the AS-Level Internet topology by merging the
snapshots from various monitors.

In our following study of network evolution, for each topological property, we will first leverage different
data sets to make a rough judgement on whether the property in question is sensitive to the number and set of
monitors, and then choose the appropriate data set for further reasoning. Taking the IPv4 network as an
example, for those properties that are insensitive to the number of monitors or can be gracefully
characterised by existing monitors, we will focus on the OIX data set to supply a comprehensive view of the
evolutionary trends of these properties. While for those properties that are sensitive to monitors, we will
primarily rely on the Set52 data set to make our conclusions and use other data set conservatively, hoping to
minimise the effect of biased sampling due to monitor variation.






\begin{figure*}[htb]
\centering \subfigure[IPv4] {
\includegraphics[width=8cm]{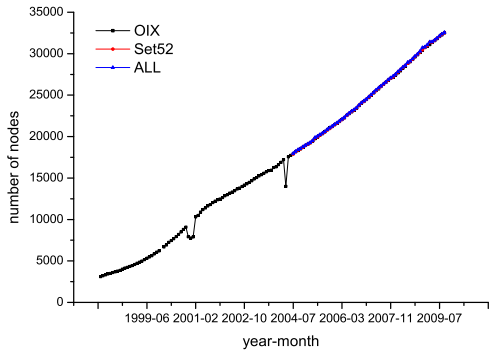}\label{ipv4-node}
} \subfigure[IPv6] {
\includegraphics[width=8cm]{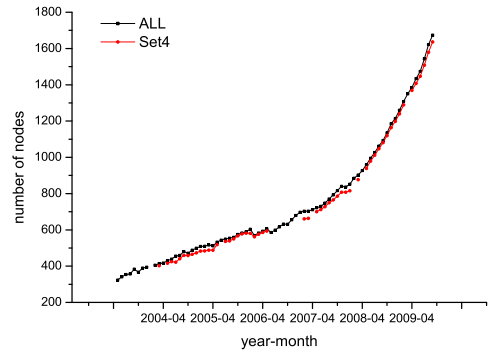}
 \label{ipv6-node}
}
 \subfigure[Curve fittings for IPv4 (OIX) ]{
\includegraphics[width=8cm]{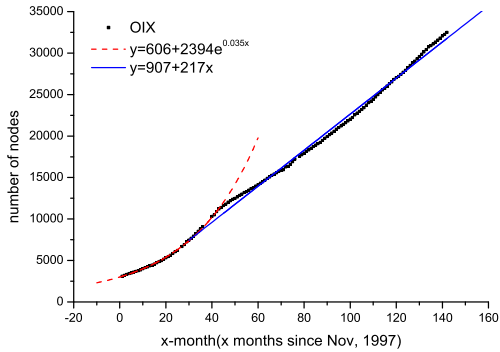}\label{ipv4-node-fit}
} \subfigure[Curve fittings for IPv6 (ALL)] {
\includegraphics[width=8cm]{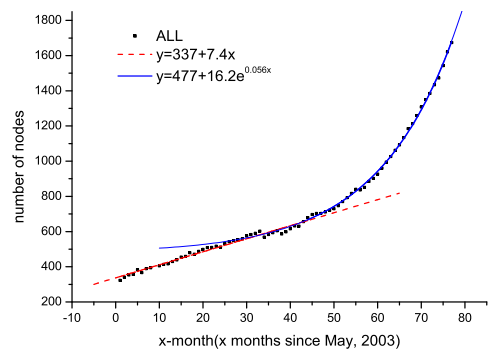}
 \label{ipv6-node-fit}
 }
\caption{Evolution of the network size (number of nodes) of IPv4 and IPv6 networks.} \label{node}
\end{figure*}

\begin{figure*}[htb]
\centering \subfigure[IPv4]{
\includegraphics[width=8cm]{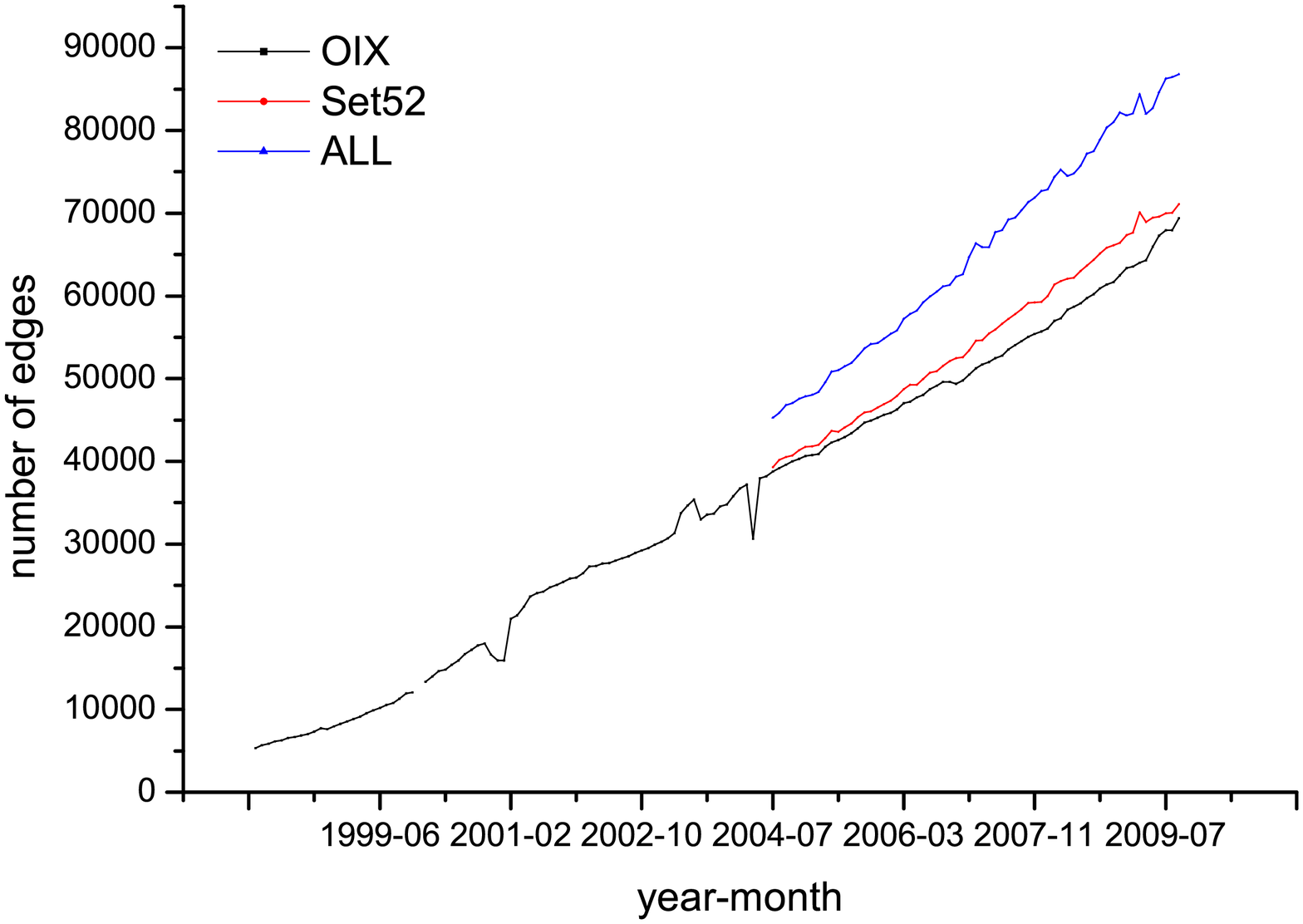}\label{ipv4-edge}
} \subfigure[IPv6]{
\includegraphics[width=8cm]{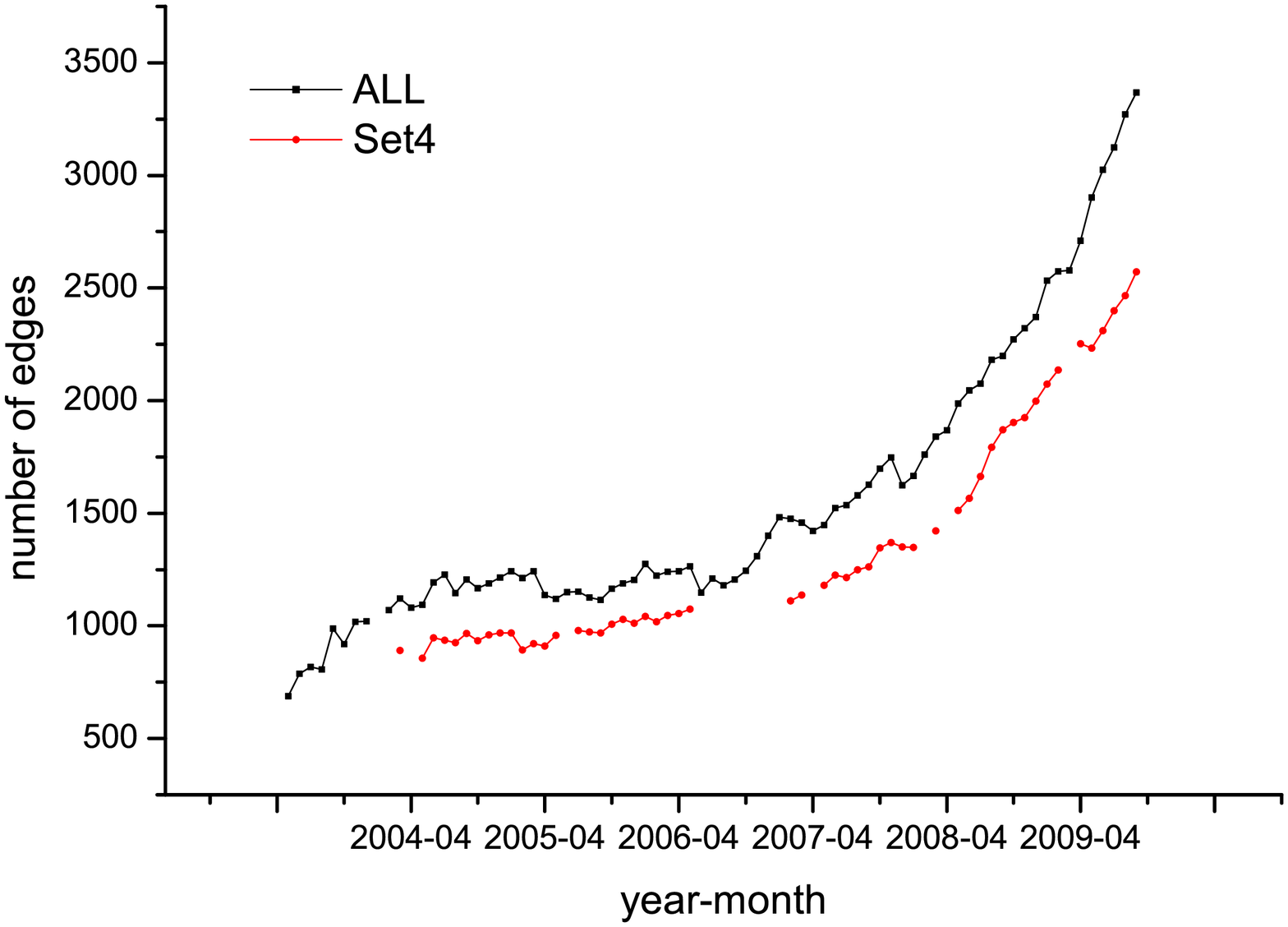}
 \label{ipv6-edge}}
\caption{Evolution of the number of edges of the IPv4 and IPv6 networks.} \label{edge}
\end{figure*}

\section{Evolution of the IPv4 and IPv6 AS-Level Internet
Topologies}\label{evolution-study}

Here we perform a side-by-side evolutionary study of several
important graph properties of  the IPv4 and IPv6 AS-Level
topologies. These properties include network size (number of nodes
and edges), degree properties (maximum degree, average degree, and
degree distribution), average shortest path length, clustering
coefficient and assortative coefficient.

\subsection{Network Size}

The first and foremost question of network evolution is how the network size evolves over time. Network size
consists of two aspects: the number of nodes and the number of edges. However, the limited number of BGP
monitors has significant impact on the number of edges that can be discovered. It has long been recognised
that the AS topology inferred from BGP data will systematically lose a large fraction of peer-to-peer links
\cite{three-data-sources,internet-dark-matter,a-framework-for-discovering-missing-links,in-search-of-the-elusive-ground-truth,impacts-sampling-bias}.
Nevertheless, the monitor issue almost has no effect on the number of nodes that can be detected
\cite{three-data-sources,in-search-of-the-elusive-ground-truth,impacts-sampling-bias,route-monitor-selection}.
These perceptions are obviously confirmed by Fig.\,\ref{node} and Fig.\,\ref{edge}, from which we observe
that whatever the data set is, the number of nodes observed are similar, albeit the number of detected edges
can show significant differences.

Since the number of monitors has little effect on the number of ASes that can be detected, we can rely on the
OIX data set to make a long time study of the evolution of the number of nodes over the past 12 years. It is
easy to find that from 1997 to 2001, the number of nodes obeyed an exponential growth rate, but after that,
it can be better described by a linear growth process. This effect has also been reported in
\cite{ten-years-evolution}.

The IPv6 network is different from the IPv4 network. In the early days from 2003 to 2006, the number of nodes
grew linearly, which was also reported in \cite{basic-property-ipv6}. However, after 2006, the number of
nodes grows exponentially. Fig.\,\ref{ipv4-node-fit} and Fig.\,\ref{ipv6-node-fit} present the fitting
functions of the two curves of IPv4's OIX data set and IPv6's ALL data set (we excluded some apparently
exceptional points in the OIX data set during the fitting, and renumbered the months by sequential numeric
numbers). The result is that, in the IPv4 network, the leading portion of the curve grows exponentially with
$y\sim e^{0.035x}$, and the rest grows linearly with $y\sim 217x$. While in the IPv6 network, the leading
portion grows linearly with $y\sim 7.4x$, and the rest grows exponentially with $y\sim e^{0.056x}$. It is
interesting to observe that the exponential growth rate of IPv6 after 2006 is even faster than the
exponential growth rate of IPv4 before 2001.


The number of edges has similar growth trends, however, since the
number and set of monitors vary over time, any conclusions made on
the edges should be taken cautiously.

The difference between the growth patterns in the number of ASes in
the IPv4 and IPv6 networks is an indication of the different
development stages of these two networks. The IPv4 network, after a
rapid exponential growth, enters into a more stable stage, whereas
the IPv6 network is still in the exponential growth stage.

\begin{figure*}[htb]
\centering \subfigure[IPv4]{
\includegraphics[width=8cm]{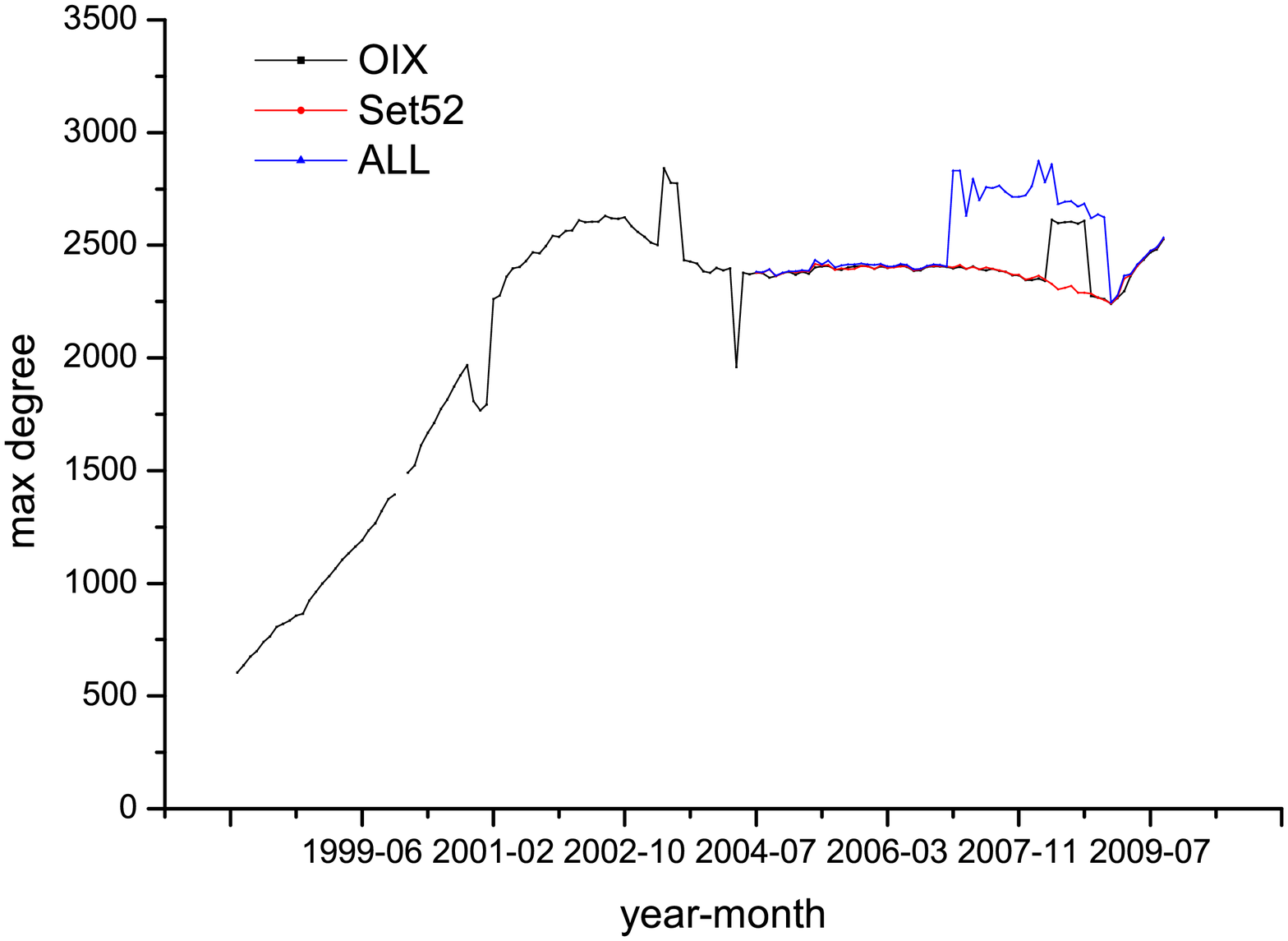}\label{ipv4-max-degree}
} \subfigure[IPv6]{
\includegraphics[width=8cm]{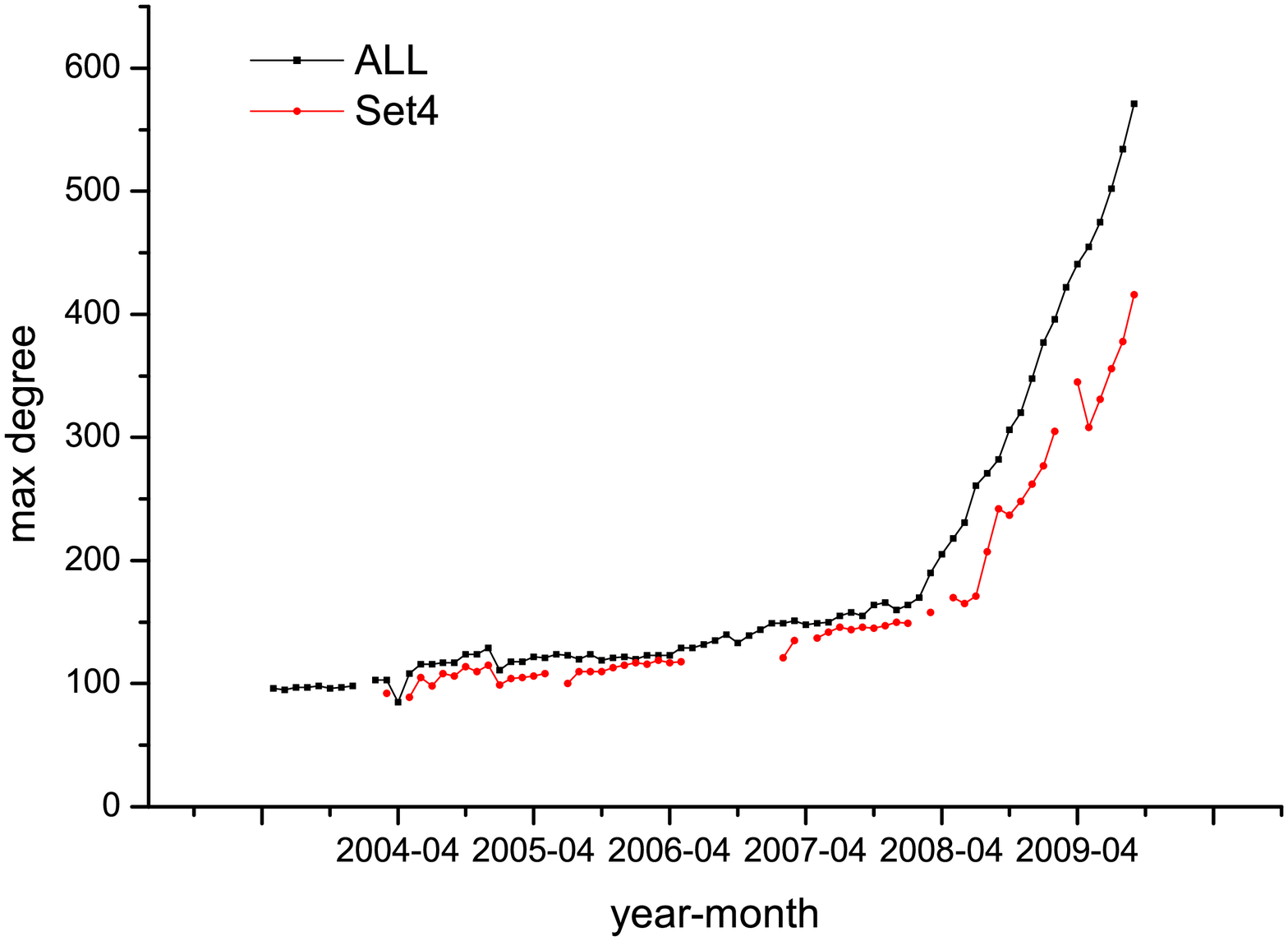}
 \label{ipv6-max-degree}}
\caption{Evolution of the maximum degree of the IPv4 and IPv6 networks.} \label{max-degree}
\end{figure*}

\subsection{Maximum Degree}
Fig.\,\ref{max-degree} reports the evolution of maximum degrees in the two networks. The maximum degree is a
particularly important topological property in the AS topology because it is often far larger than what the
typical preferential attachment models would predict and hence plays a crucial role in ensuring the network
connectivity. It is observed that despite the different sets of monitors used to construct the graphs, the
growth trends of maximum degree are similar. This means the maximum degree is largely unaffected by the
limitation of the current monitoring system. In fact, the nodes with the highest degrees are always the
tier-1 transit ASes. As is shown in \cite{in-search-of-the-elusive-ground-truth}, the current public view of
BGP monitors are sufficient to detect all the neighbouring links of these tier-1 ASes in the IPv4 network, so
the maximum degree is largely unaffected by deploying more monitors. In comparison, there is still a slight
gap of the maximum degree between Set4 and ALL data set in the IPv6 network, which implies that 4 monitors
are insufficient to capture all the neighbouring links of tier-1 ASes in the IPv6 network.

In the IPv4 network, the maximum degree grew rapidly from 1997 to 2001, after which it remained relatively
stable. Our previous analysis on the data from Dec, 2001 to Dec, 2006 also confirmed that the maximum degree
of IPv4 AS-level topology remains quite stable \cite{evolution-of-internet-and-core}. In this paper, we give
a more comprehensive picture of its evolution over a much wider temporal spectrum. In the IPv6 network, the
maximum degree grew slowly from 2003 to 2007(similar result was also reported in \cite{basic-property-ipv6}),
while after that it entered into a rapid growth stage. The maximum degree growth pattern is another
indication that the IPv6 network is currently in the rapid expansion stage.

\begin{figure*}[htb]
\centering \subfigure[IPv4]{
\includegraphics[width=8cm]{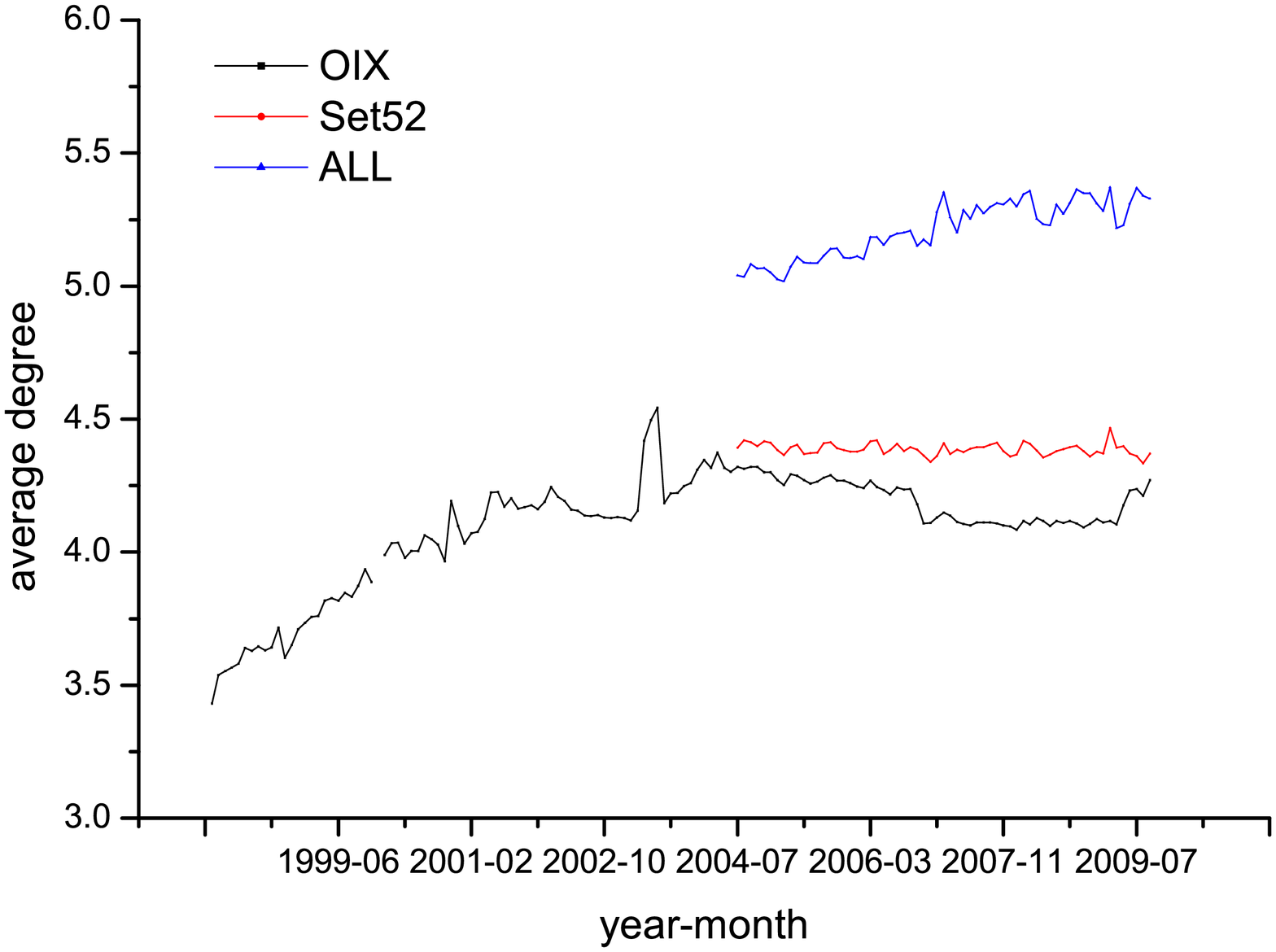}\label{ipv4-average-degree}
} \subfigure[IPv6]{
\includegraphics[width=8cm]{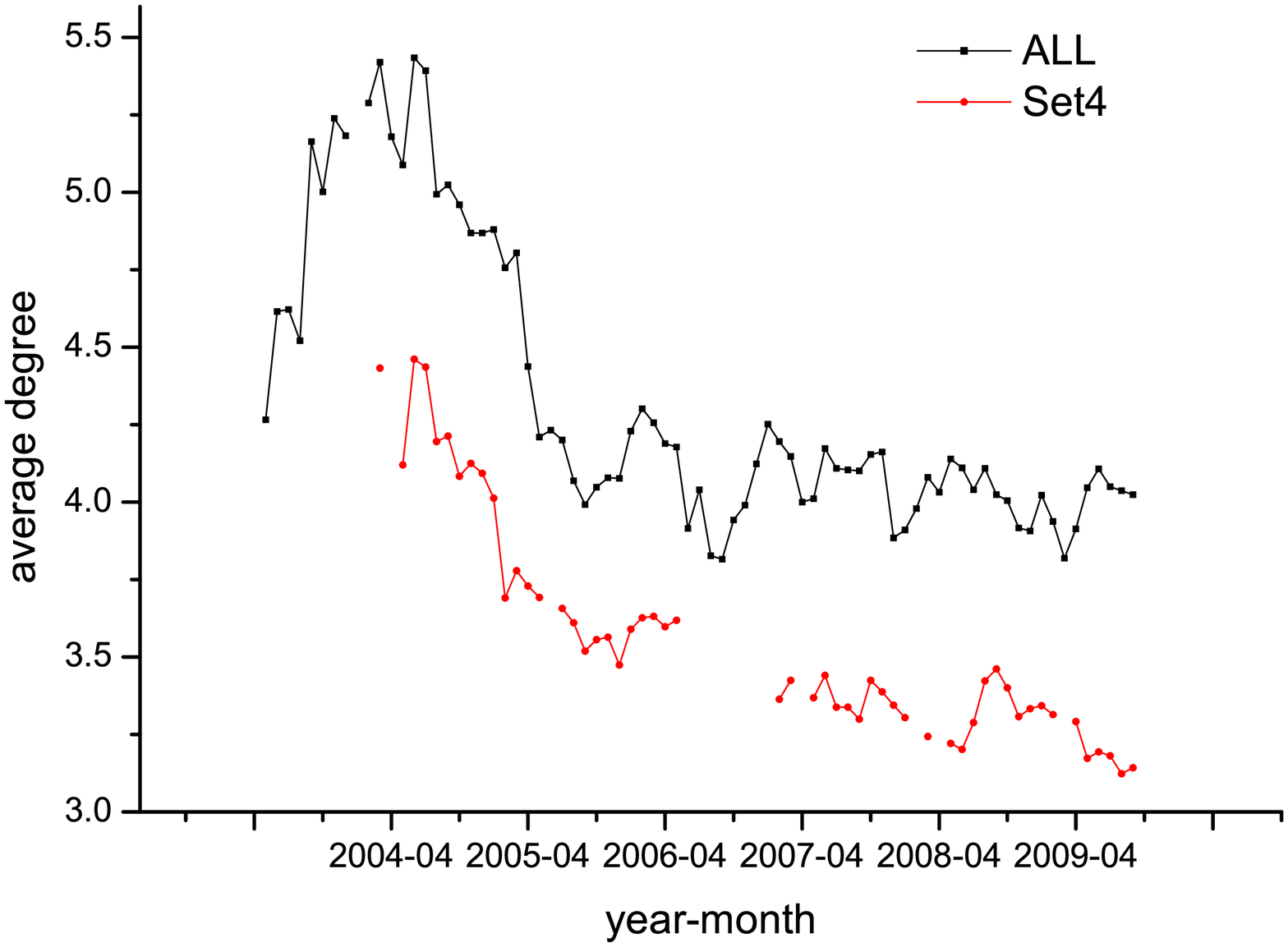}
 \label{ipv6-average-degree}}
\caption{Evolution of the average degree of the IPv4 and IPv6 networks.} \label{average-degree}
\end{figure*}

\subsection{Average Degree}


The density of connectivity in a network can be indicated by the average degree of nodes, which can be given
as $2L/N$ where $L$ is the number of edges and $N$ is the number of nodes.
Fig.\,\ref{ipv4-average-degree} shows the evolution of average degree of the IPv4 and IPv6 networks. For the
IPv4-OIX data set, the average degree was increasing until 2001 and then it remained relatively stable. For
the IPv4-Set52 data set, the average degree is also very stable in recent years.
These observations do not support previous claims that the AS-Level
Internet topology was a so-called accelerating
network~\cite{effect-of-accelerating-growth,accelerating-networks},
or the Internet followed the so-called densification
law~\cite{densification-laws,graph-mining}, i.e.~the number of edges
grows faster than the number of nodes, or equivalently, the average
node degree increases. This claim may be correct before the phase
change, but it stops accelerating after the phase change.
%
For the IPv4-ALL data set, the average degree was much larger and was still increasing. The larger average
degree is due to the larger number of monitors, but it is not clear whether the still increasing average
degree is also due to  the increasing number of monitors.
%
%
Nevertheless, the average degree of the IPv6 network, however, exhibits an remarkably different evolutionary
trend, where the average degree is in fact decreasing rapidly in recent years. This is the trend for both the
IPv6-Set4 and IPv6-ALL data sets.

\begin{figure*}[htb]
\centering \subfigure[IPv4]{
\includegraphics[width=8cm]{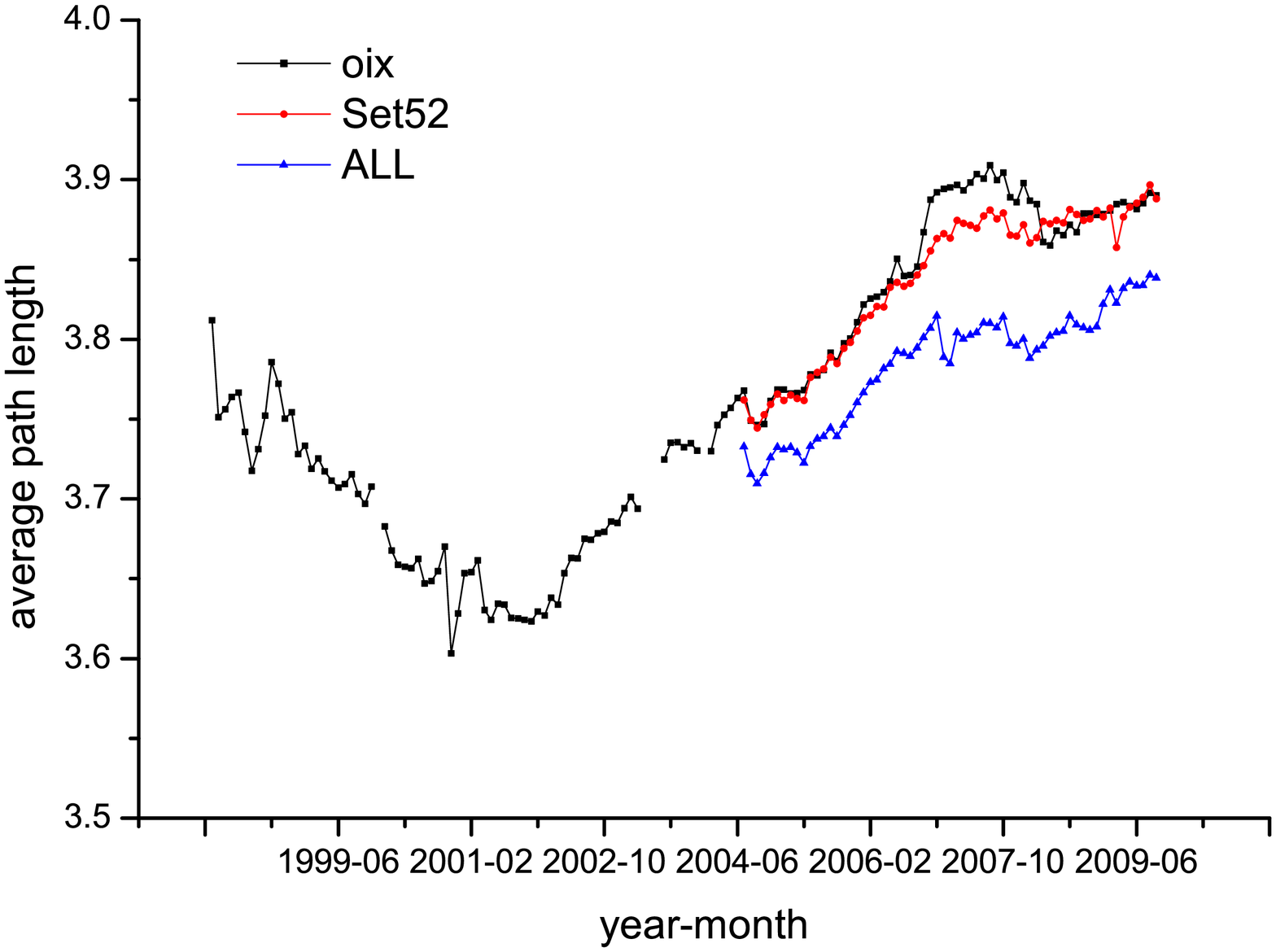}\label{ipv4-apl}
} \subfigure[IPv6]{\includegraphics[width=8cm]{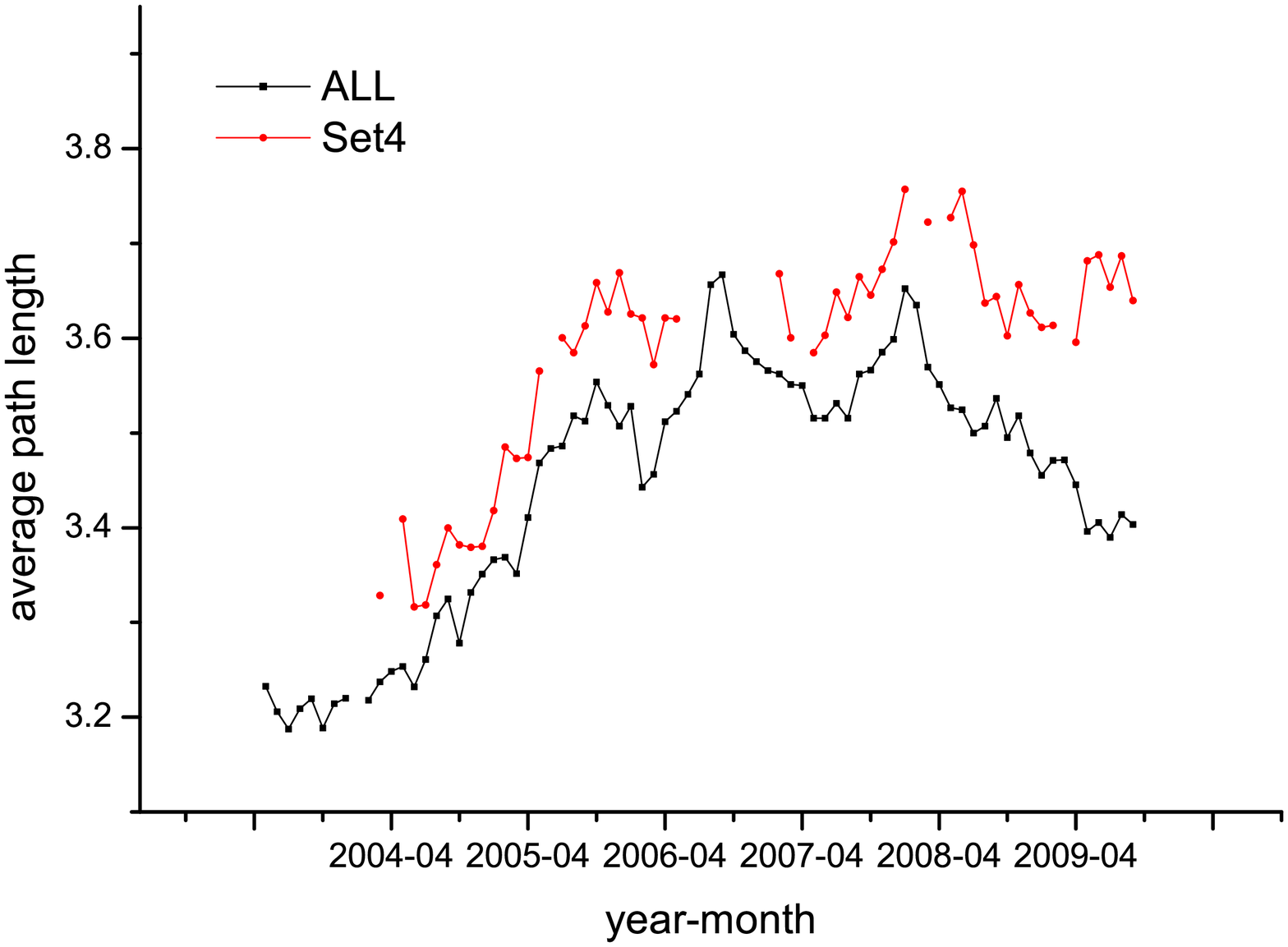}\label{ipv6-apl}} \caption{Evolution of the average
shortest path length of the IPv4 and IPv6 networks.} \label{apl}
\end{figure*}

\subsection{Shortest Path Length}

A key topological property of a network is the  average shortest
path length between any pair of nodes, which indicates a network's
routing efficiency if all traffic follows the shortest path
available\footnote{In reality, inter-AS paths are also constrained
by the routing policies~\cite{as-relationship} and real AS paths may
be inflated compared with the shortest
paths~\cite{AS-path-inflation}, but minimising the number of hops is
still a major criterion in path selection and the shortest path also
reflects the best achievable routing efficiency.}.

It is reported in~\cite{densification-laws} that the average
shortest path length of the IPv4 AS graph was shrinking, whereas  it
was reported in~\cite{evolution-of-internet-and-core} that this
measure was increasing. This contradiction arises from the fact that
these two works investigated different stages of the Internet
evolution. In the former, the data was collected from 1999 to 2001,
just before the phase change; while in the latter, the data was from
2001 to 2006, just after the phase change.

We show in Fig.\,\ref{ipv4-apl}  the  average shortest path length of the IPv4 network evolution over a much
longer  period encompassing the two previous works.
The OIX data set shows that following a few years of decreasing, the average shortest path started to grow in
2001. The decreasing of the average shortest path in the early years is arguable because it is unclear
whether it was merely due to the lack of monitors. Nevertheless,all three data sets show that the average
shortest path length of the IPv4 network is increasing in recent years.

Taking this phenomenon with the evolutionary trend of the maximum degree, we can conjecture that before 2001,
the IPv4 network was at an evolution stage when there was a boom of newly born ASes. The new ASes tended to
connect to the most-connected ASes (or tier-1 ASes), and the most-connected ASes rapidly enriched their
mutual peering relationships (i.e. rich-club phenomenon~\cite{rich-club}). As a result, the maximum degree
increased rapidly, and the average shortest path length \emph{might} shrink. After 2001, the core of the IPv4
became relatively stable in terms of the number of tier-1 ASes and the edges among
them~\cite{evolution-of-internet-and-core}, and newly born ASes primarily connected to tier-2 or tier-3
regional service providers. Therefore the maximum degree stopped increasing, and the average shortest path
length started to increase.

\begin{figure*}[htb]
\centering \subfigure[IPv4 (OIX)]{
\includegraphics[width=8cm]{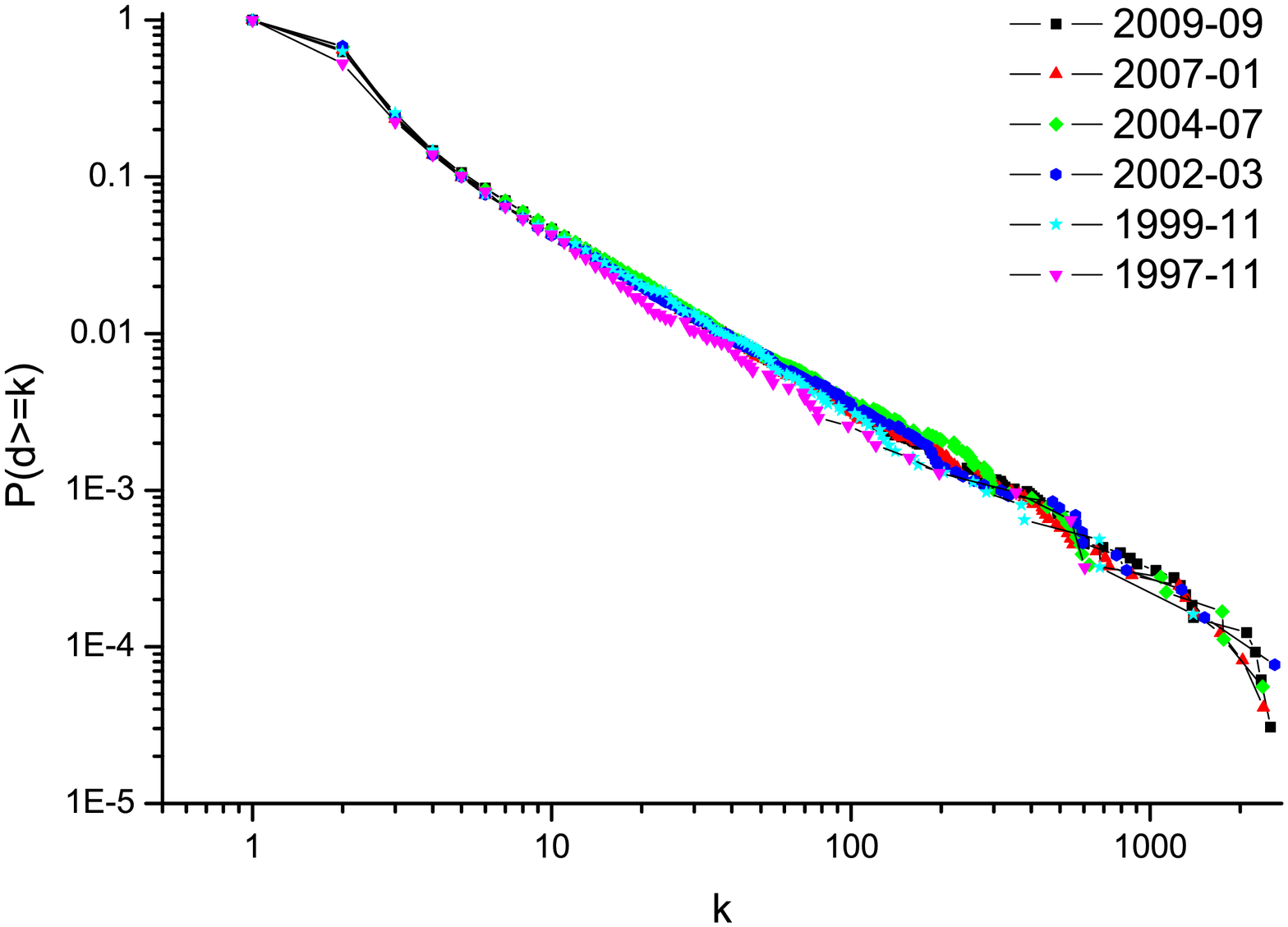}\label{ipv4-degree-distribution}
} \subfigure[IPv6 (ALL)]{
\includegraphics[width=8cm]{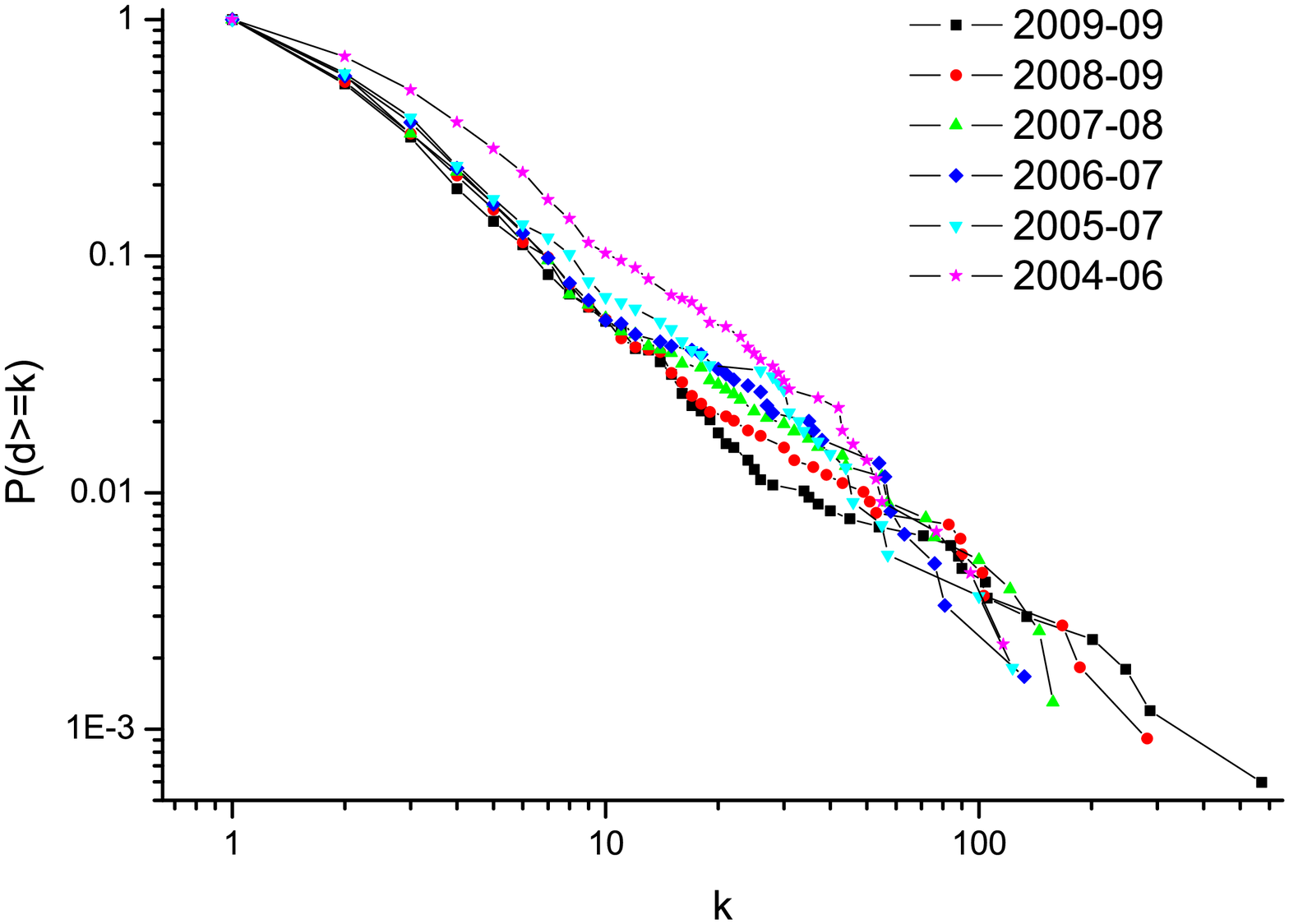}
 \label{ipv6-degree-distribution}}
\caption{Evolution of degree distributions in the IPv4 and IPv6
networks.} \label{degree-distribution}
\end{figure*}

\subsection{Degree Distribution}

Degree distribution is a frequently cited macroscopic topological
property. Fig.\,\ref{degree-distribution} shows the complementary
cumulative degree distribution (CCDF) of the IPv4 and IPv6 networks.
It is clear that the networks follow a power-law degree
distribution, $p(k)\sim k^{-r}$, as repeatedly reported
before~\cite{power-law-99, BA, chinese-as-graph, newman-review,
basic-property-ipv6, modeling-ipv6}.

For the IPv4 network, the CCDF curves of different snapshots overlap with each other with a stale power-law
exponent. For the IPv6 network, we see that the curves shift to the left as time goes with a slightly
increasing power-law exponent.

\begin{figure*}[htb]
\centering \subfigure[IPv4]{
\includegraphics[width=8cm]{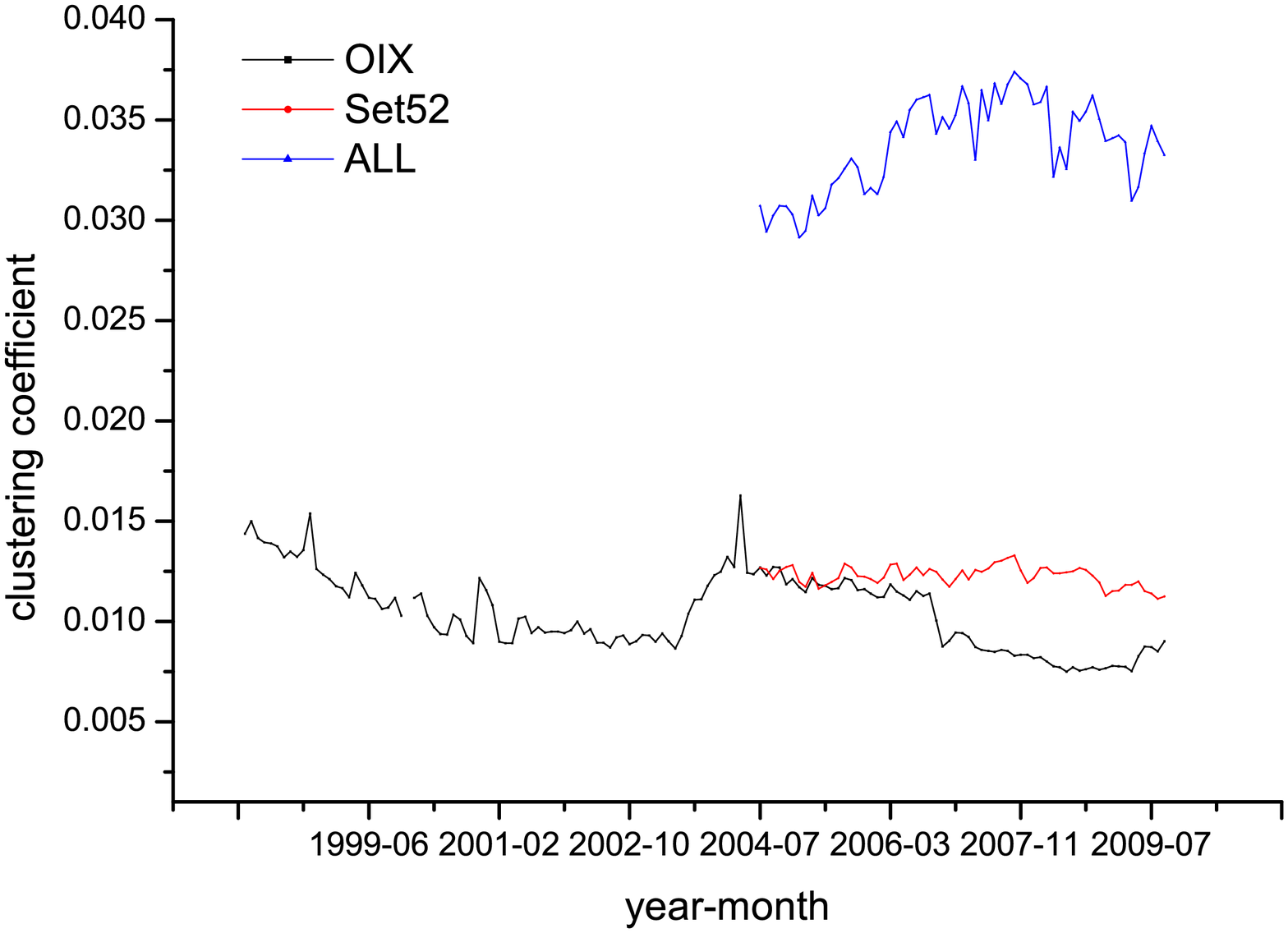}\label{ipv4-clustering}
} \subfigure[IPv6]{
\includegraphics[width=8cm]{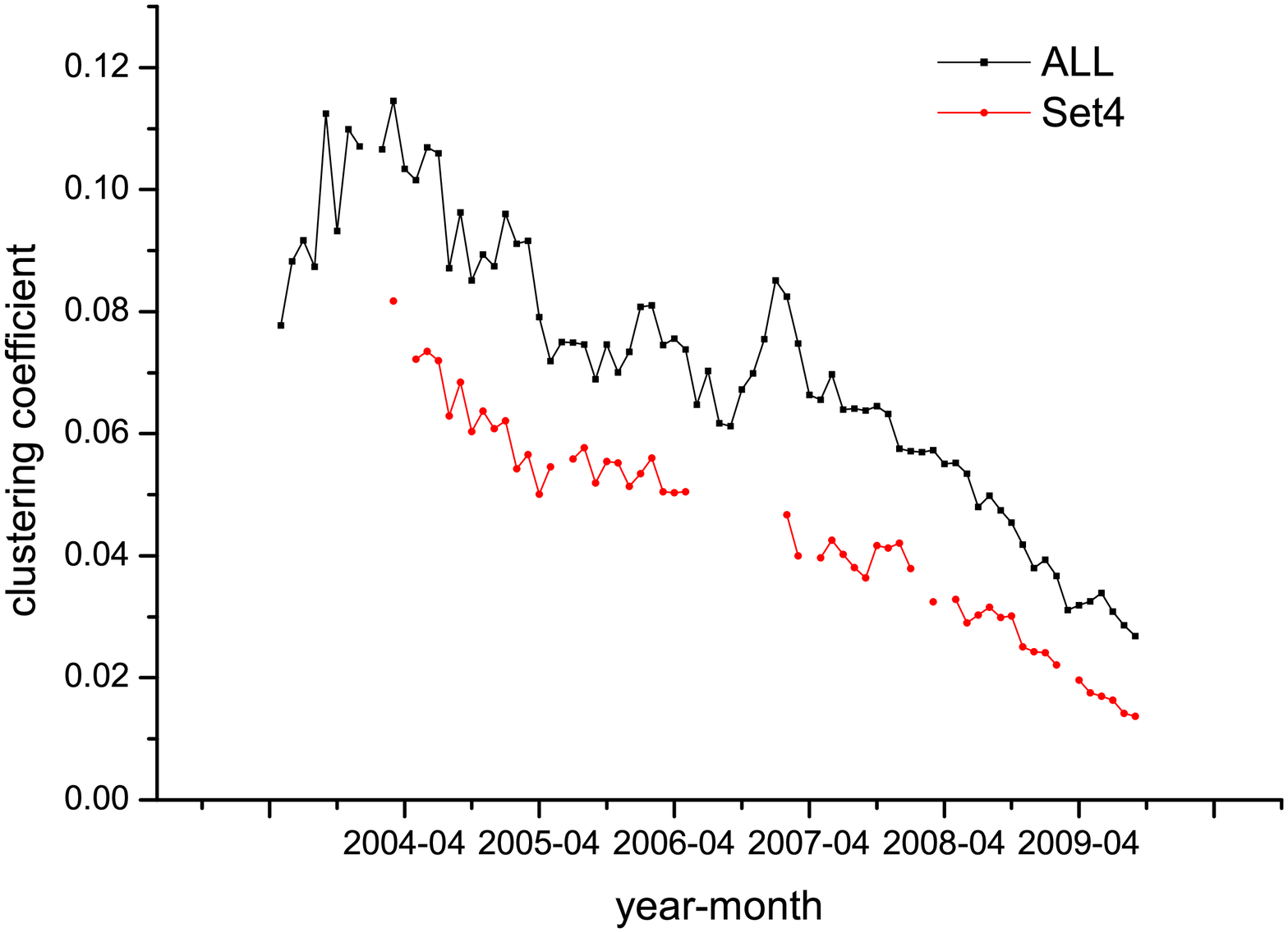}
 \label{ipv6-clustering}}
\caption{Evolution of clustering coefficients in the IPv4 and IPv6
networks.} \label{clustering}
\end{figure*}

\subsection{Clustering Coefficient}

The clustering coefficient of a network is defined as three times the ratio of the total number of triangles
to the total number of connected vertex triples in the network~\cite{clustering-correlated-networks}. It
measures the density of triangles in a network, which is relevant to alternative path and redundancy.
Fig.\,\ref{clustering} shows the evolution of clustering coefficient of the IPv4 and IPv6 networks. As
reported in \cite{in-search-of-the-elusive-ground-truth}  the number of monitors has significant impact on
the clustering coefficient. Here we discuss the IPv4-Set52 and IPv6-Set4 data sets because they maintain the
same monitors during their measurement time. The IPv4-Set52 data set shows that the clustering coefficient in
the IPv4 network is relatively stable in the past five years. In comparison, the clustering coefficient in
the IPv6 network decreases gradually in recent years.

\begin{figure*}[htb]
\centering \subfigure[IPv4]{
\includegraphics[width=8cm]{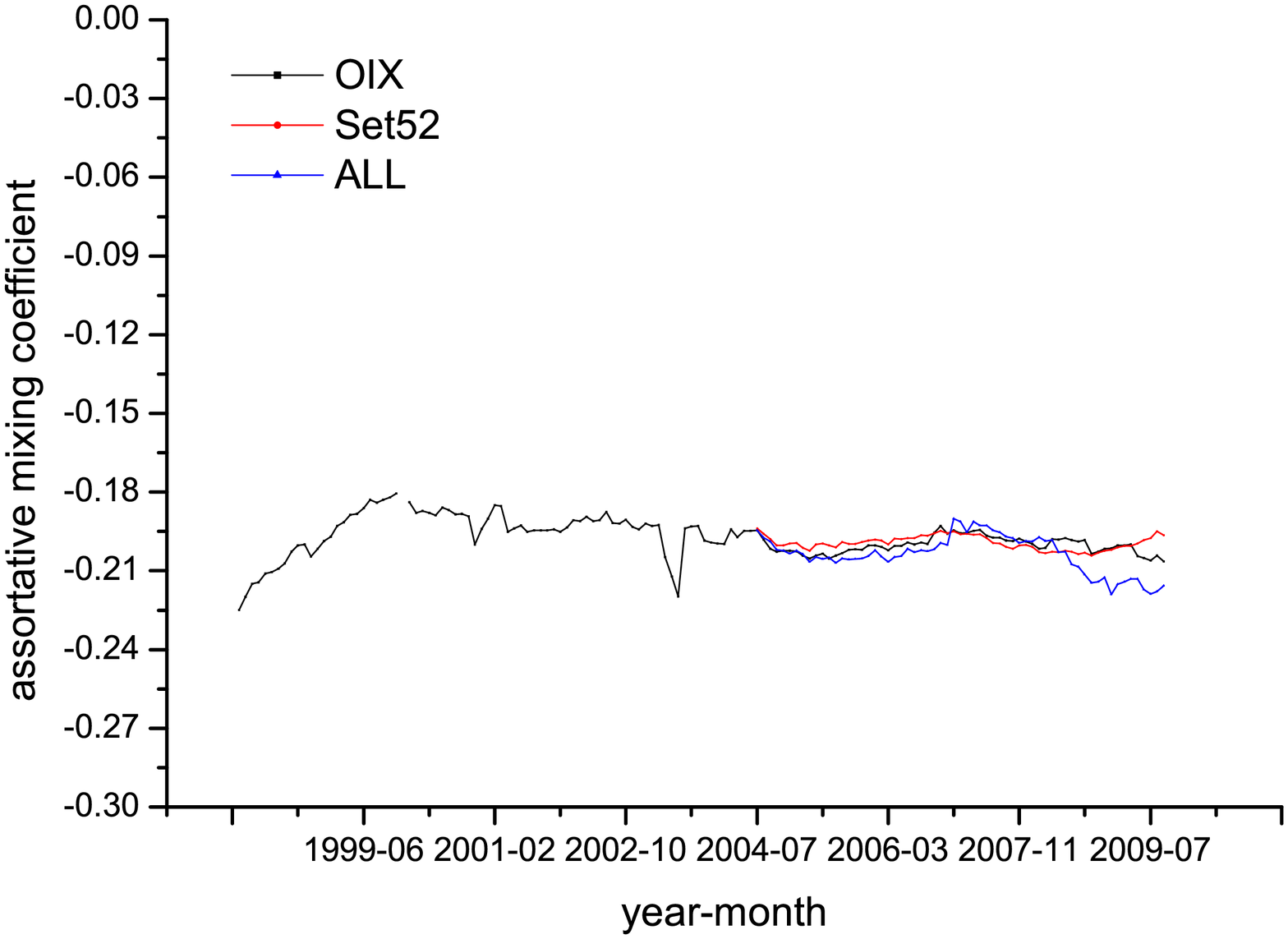}\label{ipv4-mixing}
} \subfigure[IPv6]{
\includegraphics[width=8cm]{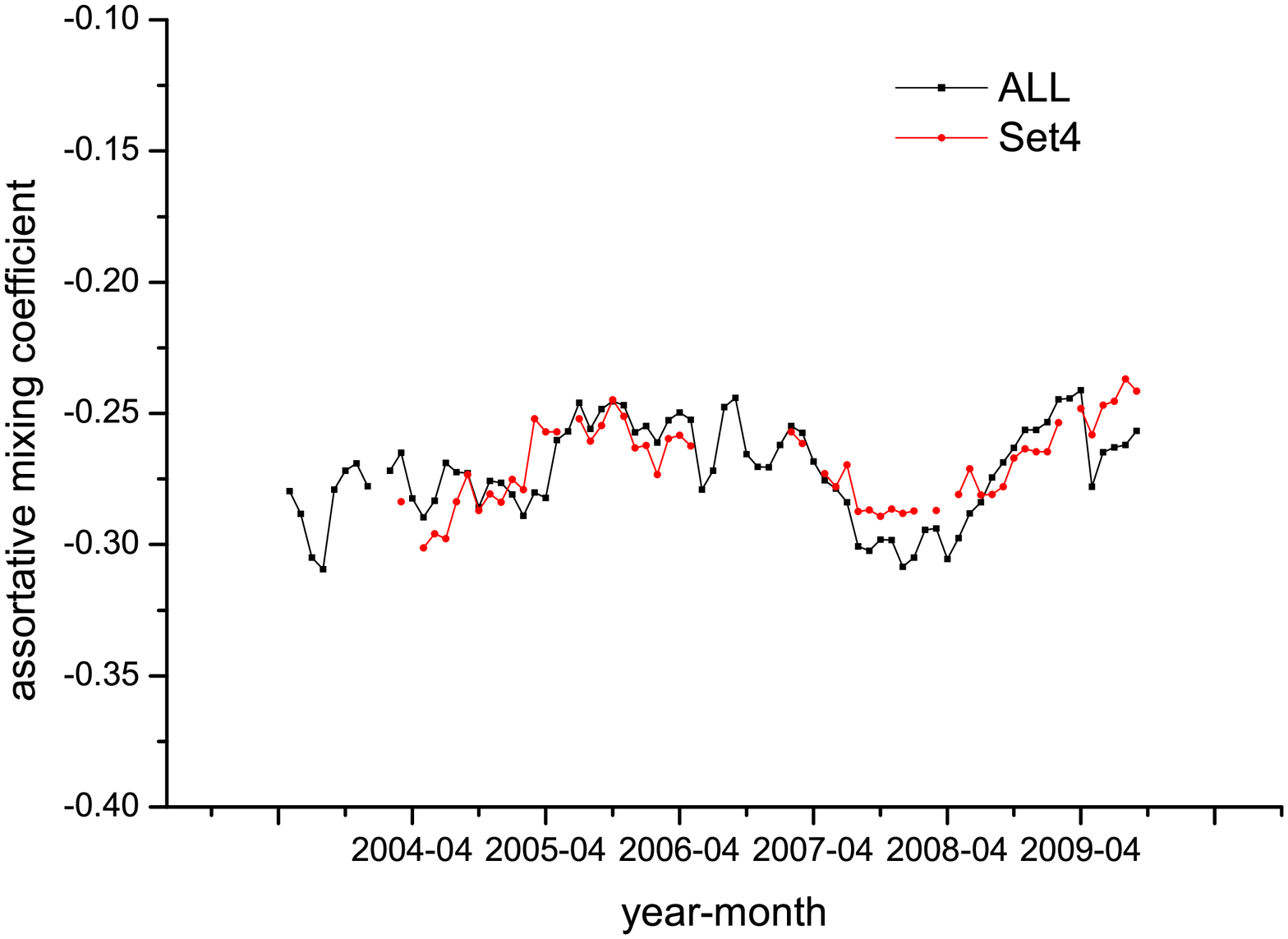}
 \label{ipv6-mixing}}
\caption{Evolution of the assortative  coefficient of the IPv4 and IPv6 networks.} \label{mixing}
\end{figure*}

\subsection{Assortative  Coefficient}

Assortative  coefficient~\cite{mixing} measures whether nodes tend
to connect to nodes of similar degrees. AS-level Internet topologies
are known to be disassortative mixing with a negative  assortative
coefficient value, i.e., high-degree nodes tend to connect to
low-degree nodes and vice versa. Fig.\,\ref{mixing} shows the
evolution of assortative coefficient of  the IPv4 and IPv6 networks.
We can see that both networks are disassortative mixing. The IPv6
network is more disassortative than the IPv4 network.
An interesting observation is that the assortative  coefficient of each of the two networks remains
relatively stable over all time, and the variation in monitors has little influence on this metric. Hence,
disassortative mixing could be viewed as an invariant for the AS-level Internet topology.

\begin{table*}
\caption{\label{tab:evolution} Summary of the Internet evolutionary phase changes}
\renewcommand{\tabcolsep}{1.0pc}
\begin{tabular*}{\textwidth}{l|cc|cc}
\hline\hline
  & IPv4 before 2001  & IPv4 after 2001 &  IPv6 before 2006 & IPv6 after 2006\\
\hline
Number of nodes & exponential growth & linear growth  & linear growth & exponential growth\\
Number of edges & exponential growth & linear growth  & slow growth & rapid growth\\
%
%
Maximum degree  & rapid growth &  stable  & slow growth & rapid growth\\
Average degree  & steady growth &  stable   & rapid decreasing & slow decreasing \\
Shortest path length  & decreasing & increasing  & increasing &  stable \\

\hline
Degree distribution  & \multicolumn{2}{c|}{power-law} & \multicolumn{2}{c}{power-law} \\
Power-law exponent & \multicolumn{2}{c|}{stable} & \multicolumn{2}{c}{slightly increasing} \\
Clustering coef.  & \multicolumn{2}{c|}{ stable} & \multicolumn{2}{c}{decreasing} \\
Assortative coef.  & \multicolumn{2}{c|}{ stable} & \multicolumn{2}{c}{ stable} \\
\hline\hline
\end{tabular*}
\end{table*}

\section{Discussions\label{impacts}}

\subsection{Internet Evolutionary Phase Changes}

We summarise the above results in Table~\ref{tab:evolution}. It is clear that both the IPv4 and the IPv6
networks have experienced an evolutionary phase change. The phase change of the two networks, however,
happened at different times with different transition patterns.

For the IPv4 network, the phase transition took place around year 2001 when the network changed from a
process of rapid growth to a stage of slow growth with relatively stable structure. One possible reason could
be the burst of the dot-com bubble at the beginning of this century which slowed down the investment on
Internet. There might be technical reasons as well, such as the near exhaustion of AS numbers and the
increasing size of BGP routing tables.

For the IPv6 network, the phase transition took place in year 2006 when the network changed from a stage of
relatively slow growth to a process of rapid expansion. This may relate to a number of events happened around
that time, including a boom of IPv6 deployment projects around the world, such as the CNGI project in China,
and the plan to phaseout  the 6bone, which is an IPv6 network that extensively relied on the IPv6-over-IPv4
tunnelling technique~\cite{6bone-phase-out}. The exact reasons for the phase changes will be investigated in
our future work.

\subsection{IPv6-over-IPv4 Tunnelling}

\begin{figure*}[htb]
\centering \subfigure[]{
\includegraphics[width=8cm]{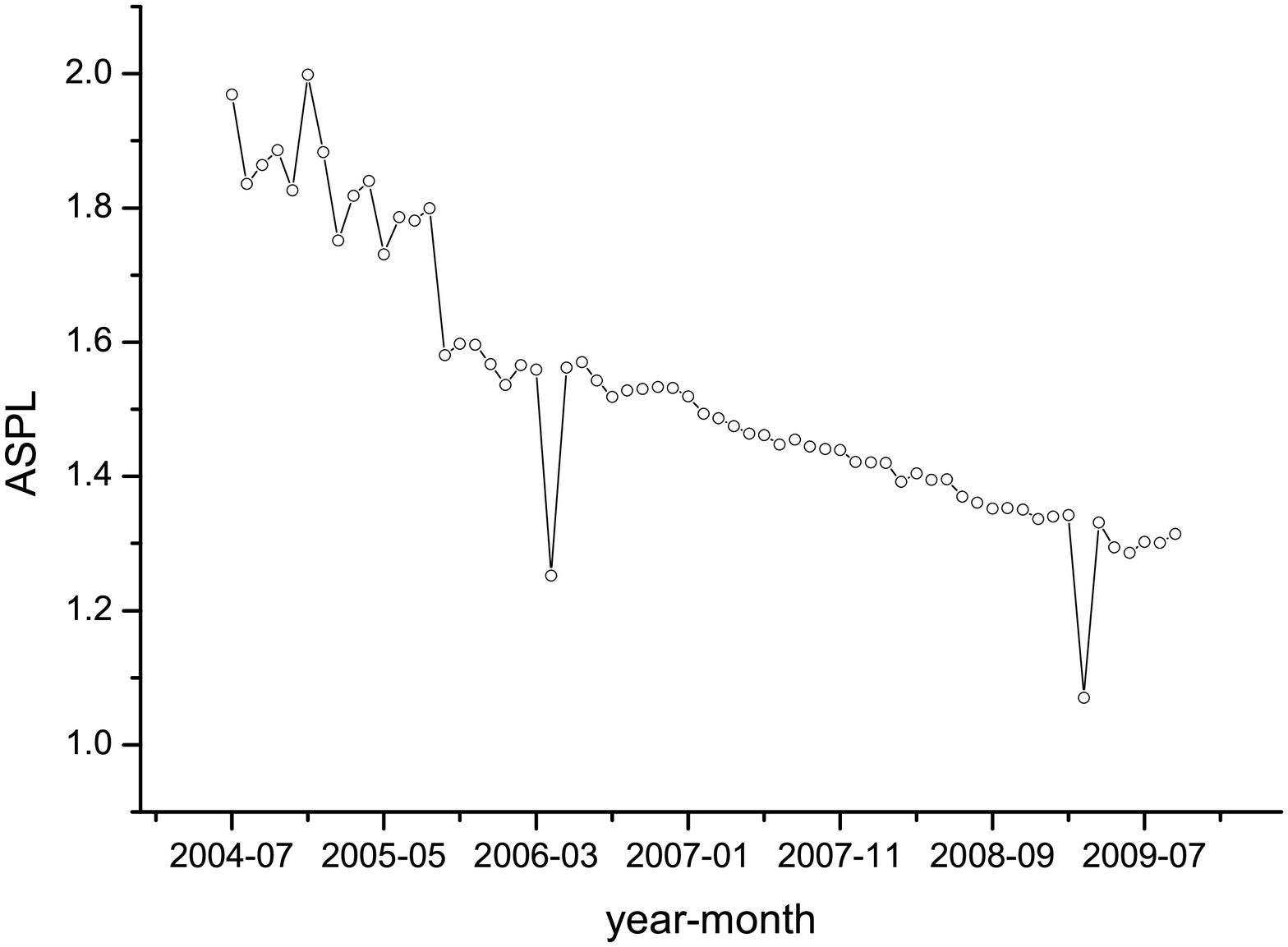}\label{aspl-tunneling}
} \subfigure[]{
\includegraphics[width=8cm]{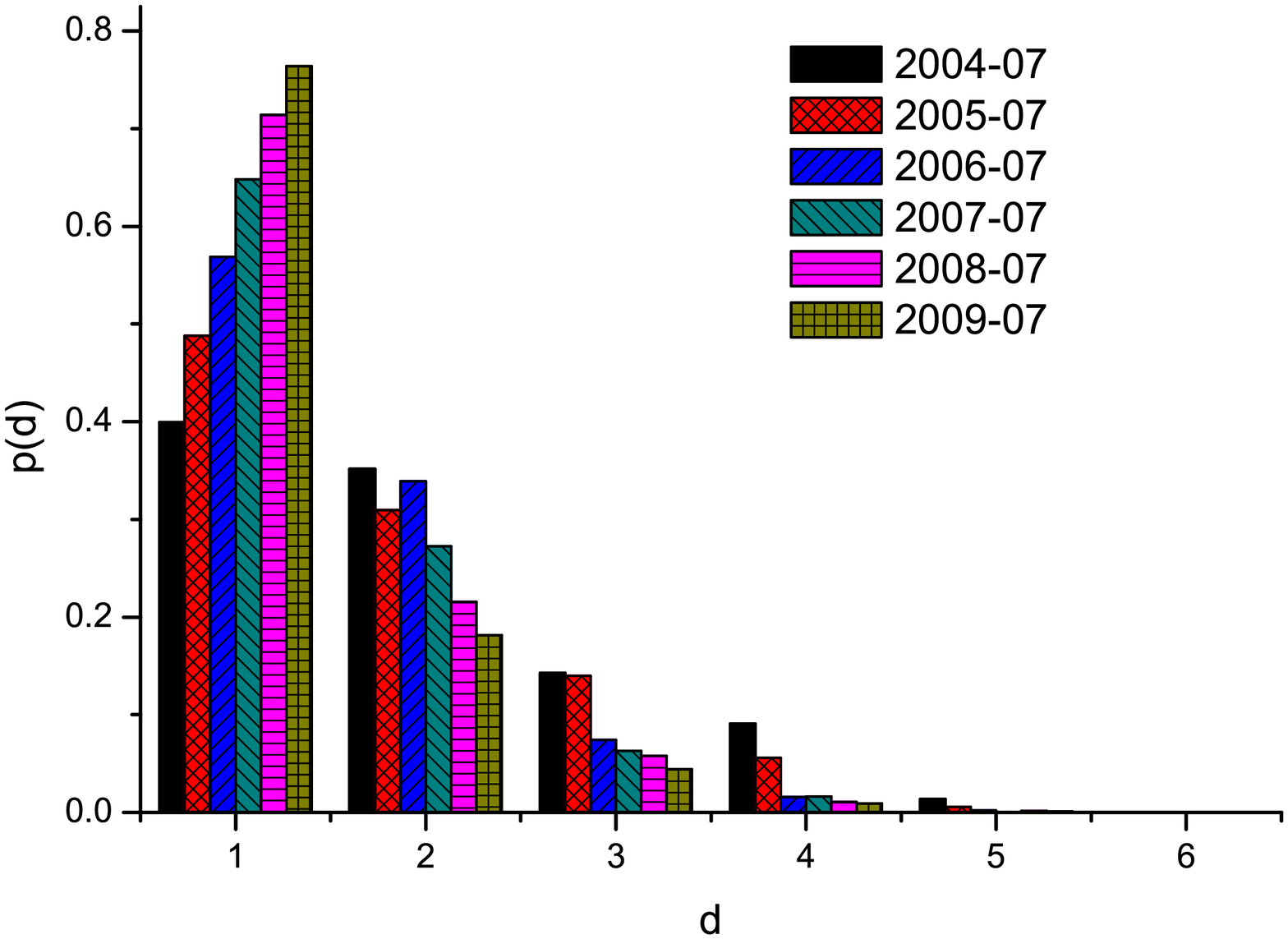}\label{tunneling-distribution}
}
\caption{The impact of IPv6-over-IPv4 tunnelling on the Internet evolution. (a) Evolution of the average
shortest path length (ASPL) of the connected IPv6 AS pairs on the IPv4 network. (b) Evolution of the
distribution of shortest path length of the connected IPv6 AS pairs on the IPv4 network, where $p(d)$ is the
probability that a connected pair of ASes on the IPv6 network has path length $d$ on the corresponding IPv4
network. }\label{tunnel}
\end{figure*}

One of the fundamental differences between the evolution of IPv6 and
IPv4 is that the growth of the IPv4 topology followed the growth of
the physical infrastructure of the Internet, whereas the growth of
the IPv6 topology was more a matter of deploying the IPv6 technology
over the existing (and changing) infrastructure. On the other hand,
for the same reason the two networks are also related. The
IPv6-over-IPv4 tunnelling, for example,  has been widely used in the
early years of IPv6 deployment. This technique allows two ASes on
the IPv6 network appear to be directly connected with each other
whereas in fact there might be a number of hops between them on the
underlying IPv4 network. IPv6 network is more disassortative than
the IPv4 network (see Fig.\,\ref{ipv6-mixing}), which may partly
arise from the tunnelling deployment of IPv6.

To  study the impacts of the tunnelling technique on the evolution
of the IPv6 network, we use the following approach: for an edge
(AS1, AS2) on an IPv6 snapshot,  if AS1 and AS2 are also present in
the corresponding IPv4 snapshot, we compute the shortest path length
between AS1 and AS2 on the IPv4 snapshot. Intuitively, if
IPv6-over-IPv4 tunnelling is prevalently used, then the shortest
path length between AS1 and AS2 on the IPv4 snapshot will have high
probability to exceed 1. We plot the evolution of the average
shortest path length (ASPL) as well as the distribution of path
length, $d$,  for all such AS pairs in Fig.\,\ref{aspl-tunneling}
and Fig.\,\ref{tunneling-distribution} respectively. It can be seen
that the ASPL is indeed well above 1 and it decreases as time goes
by. Also the probability that a connected pair of ASes on IPv6 are
also directly connected on the IPv4 network (i.e.~the columns with
$d=1$) increases over time. This suggests that the IPv6 gradually
shifts from the tunnelling phase to the genuine IPv6 connectivity
phase. This is in accordance with the IETF's phaseout planning of
the 6bone in 2004~\cite{6bone-phase-out}. We expect this trend
towards the deployment of genuine IPv6 sessions will bring a
diminishing difference between the length of IPv6 and IPv4 paths.

%
%

\subsection{Internet Measurement Monitors}

As any work based on Internet measurement data, our work would still
be affected by the limited number of vantage points. All along the
paper, we try however as much as possible to show how the number of
monitors affects each metric. In general we note that data from one
collector is insufficient for the IPv4 network. Set52 and ALL data
sets often capture very similar network structures although the ALL
data set contains significantly more monitors. For the IPv6 network,
the ALL data set is more appropriate than the Set4 data set.

\begin{figure}[htb]
\centering
\includegraphics[width=8cm]{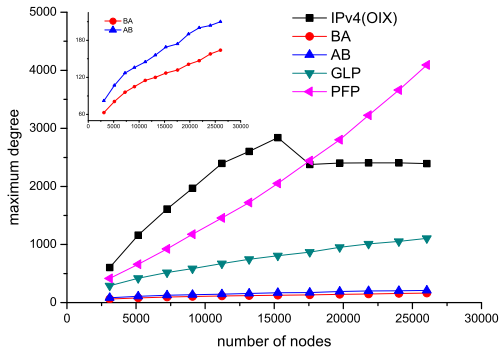}
\caption{Evolution of maximum degree of real AS topology(OIX) and four typical generative models. BA network
is generated with $m$=2 \cite{BA}. AB network is generated with $p$=0.2, $q$=0.3 and $m$=2 \cite{AB}. GLP
network is generated with $p$=0.4695 and $\beta$=0.6447 \cite{GLP}. PFP network is generated with $p$=0.4 and
$\delta$=0.021 \cite{understanding-the-evolution-of-internet}. Evolution of maximum degrees in BA and AB is
further illustrated in the inset since the growth trends are not legible in the original plot due to the
relatively small absolute values.}\label{model-maximum-degree}
\end{figure}

\subsection{Generative AS-level Internet Models}

A number of generative models have been proposed to reproduce and
explain the evolution of networks. These models are different in
many ways but they are all based on a common assumption that a
network obeys a non-changing, uniform growth mechanism throughout
its evolution. This, however, is clearly not the case for the IPv4
and IPv6 networks.

Taking the maximum degree as an example,
Fig.\,\ref{model-maximum-degree} shows that the maximum degree grows
monotonically with the number of nodes for each of the four typical
AS-Level generative models, whereas the maximum degree of the IPv4
network (OIX data set) exhibits an evolutionary phase change where
after a critical point the maximum degree became relatively stable.
This is not surprisingly as the models were not designed to
reproduce such phase change.

Our work highlights that it is not sufficient to validate a
generative Internet model against a few snapshots of the network.
Rather, we should  validate a model against long-term  evolution
data. Our results on the evolutionary phase changes of the Internet
networks provide new input for designing and validating future
Internet models.

Indeed, pure graph-theory based generative models have already been
questioned in both the router-level and AS-level
topologies~\cite{igen,HOT,optimization-driven-model,to-peer-or-not-to-peer,
modeling-internet-dynamics}. It has been recognized that the
router-level topology can be more accurately modeled by
optimization-driven approaches, e.g., HOT~\cite{HOT} and
IGen~\cite{igen}, and could be designed in a cost-effective
manner~\cite{cost-effective-network-disigning}. However, these
challenges are not raised from the perspective of the phase change.
A possible direction for future research could be to borrow some
ideas from the optimization-driven approaches to the development of
AS-Level generative models that capture the phase change.

\section{Conclusion}\label{conclusion}

In this paper, we performed an in-depth side-by-side study of the
evolution of  the IPv4 and IPv6 Internet topologies at the
autonomous system level based on historic data over a long  period
of time. Amble evidence shows that both networks have undergone a
phase change in their evolution process.  For the IPv4 network, the
approximate phase transition occurred around 2001; while for the
IPv6 network, the phase transition took place around late 2006. The
phase transition pattern of the two networks are quite different.
While the IPv4 slowed down from a rapid growth, the IPv6 has just
engaged in a fast expansion. We also found that the IPv6-over-IPv4
tunneling deployment scheme partly affects the evolution of the IPv6
network.

Our work fundamentally changes our knowledge on the Internet
topology evolution. It provides valuable input for refining existing
network models or developing new models. It also opens interesting
questions for future work, such as the exact reason for the
evolutionary phase changes of the IPv4 and IPv6 networks and the
possibility of phase changes in the future.

\section*{Acknowledgement}
GZ is  supported by National Natural Science Foundation of China
under grant number 60673168, and the Hi-Tech Research and
Development Program of China under grant number 2008AA01Z203. SZ is
supported by The Royal Academy of Engineering and EPSRC (UK) under
grant number 10216/70.

 \end{document}